\documentclass[10.5pt]{article}

\usepackage{mathptmx}
\usepackage{graphicx}
\usepackage{times}
\usepackage{amsmath, amsthm, amssymb}
\usepackage{url}
\usepackage{hyperref}
\usepackage{algorithm}

\usepackage{lscape}

\begin{document}


\begin{center}

{\Large
{How the online social networks are used: Dialogs-based structure of \texttt{MySpace} }
}\\ 
{ Milovan \v Suvakov$^{1,2}$, Marija Mitrovi\'c$^1$, Vladimir Gligorijevi\'c$^{1}$ and Bosiljka Tadi\'c$^1$}\\
{ $^1$Department of theoretical physics; Jo\v zef Stefan Institute;
 Box 3000  SI-1001 Ljubljana Slovenia; $^2$Institute of Physics, Belgrade University, Belgrade, Serbia 
\hspace{1cm}}


\end{center}

{\bf Abstract: }
\noindent
Quantitative study of collective dynamics in online social networks is a new challenge based on the  abundance of empirical data. Conclusions, however,  may depend on  factors  as user's  psychology profiles and their reasons to use the online contacts.
In this paper we have compiled and analysed two datasets from \texttt{MySpace}. The data contain networked dialogs occurring within a specified time depth, high temporal resolution, and texts of  messages, in which the emotion valence is assessed by using SentiStrength classifier.
Performing a  comprehensive analysis we obtain three groups of results:    
 Dynamic topology of the dialogs-based networks have characteristic structure with Zipf's distribution of communities, low link reciprocity, and disassortative correlations. Overlaps supporting  ``weak-ties'' hypothesis are found to follow the laws recently conjectured for online games;
 Long-range temporal correlations and persistent fluctuations occur in  the time series  of messages carrying positive (negative) emotion;  
 Patterns of user communications have 
 dominant  positive emotion  (attractiveness) and strong impact of circadian cycles and interactivity times longer than one day.
Taken together, these results give a new insight into functioning  of  the online social networks and unveil importance of the amount of information and emotion that is communicated along the social links. 
{\bf {All data used in this study are fully anonymized. }}
\\
{\bf Keywords:} Online social networks; User communities; Patterns of emotion flow; Self-organized dynamics;

\section{Introduction\label{sec-intro}}
Web as a new social space provides   ``unbearable easiness of communication'' that may lead to new social phenomena in the online world  and affect behavior of the users \cite{nature2011,kleinberg2008,cho2009}. The  online social networks nowdays represent some of the largest social structures in the world \cite{facebook2011}.  
Apart from the structure,  huge  amount of users leave digital data about their activity, which are systematically stored at Web portals, thus providing ``social experiments'' of an unprecedented scale.  Recently analysis of such data from various  Web portals has been  performed across different science disciplines \cite{castellano2009,albert_emailnets,malmgren2009,johnson2009,amichai-hamburger2010,panzarasa2009,mitrovic2010a,szell2010}. It has been recognized that the emotions, known to drive social behavior in face-to-face contacts, also play an important role in various types of the  online interactions \cite{mitrovic2010b,szell2010,ST12,paltoglou2010,thelwall2010,mitrovic2011,dodds2011,warsaw2011}.  
However, full understanding of the mechanisms that drive online social behaviors still remains elusive.  In particular,   how individual (emotional) actions of  users in the network may lead to dynamically robust collective behaviors and how to quantify the emergent phenomena, are among the questions  of primary importance. What are characteristic features of the dynamics in the online social networks in comparison with  Blogs, Forums, Games and other forms of online communications?
Here we address these questions by compiling and analysing  the empirical data of \textit{user dialogs} from the social network \texttt{MySpace}.

In recent  psychology research quantitative study of emotion   \cite{emotions-Oxford2007,scherer2005}  
 is conducted based on  Russell's multidimensional model of affect \cite{russell1980}. Specifically, each emotion known in common life can be represented  by a set of numerical values in the corresponding multidimensional space \cite{russell1980}. Three important dimensions of emotion---valence, arousal and dominance, 
can be estimated from psycho-physiological and neuronal activity \cite{calvo-review2010}. 
The  {\it valence}  measures degree of attractiveness (positive valence) or aversiveness (negative valence) to a stimuli. Similarly, the arousal and the dominance can have a range of values corresponding to different degree of reactivity to a stimuli, and power of a reaction, respectively. 
Moreover, normative emotional rating has been developed for a large number of words \cite{bradley1999}.
Based on the  lexicon of emotional words and using  machine-learning approaches recently methods  have been developed \cite{calvo-review2010,thelwall2010,paltoglou2010b} for the effective inference of the emotion content from different types of text messages appearing in the online communications.

How the online social networks are used and who uses them? The social psychology research  begin to recognize the relationship between the personal profile, social and emotional loneliness and attitudes of the Web users in general,  and users of the social networks, in particular \cite{ryan2011,yarkoni2010,amichai-hamburger2010,cheung2011}. 
The ``friendship'' association as the framework for communications in the online social networks, is partially transferred from offline social contacts. 
Recently the  topology of \textit{friendship  network} in currently largest online social site \texttt{Facebook} has been analysed \cite{facebook2011,facebook2012}. Conclusion is that ``mostly, but not entirely, agreement on common structural network characteristics'' was found \cite{facebook2011}. Somewhat different structure was reported for  the \textit{friendship  network} of \texttt{MySpace} \cite{myspace2007}.
However, it is not mere structure of the network of  declared ``friendship'' links, but rather the \textit{dynamics of  message  exchange} along these  links
 that contain relevant information for study the emergent social phenomena, in particular in the situations when   temporal bursts of (emotional) messages occur involving many users. To our knowledge, such dynamical structures, reflecting the way how the links  in the online social networks are used, have not been researched so far.
 Therefore, developing a methodology for systematic collection and analysis of the data which contain complete information about  the dynamics of dialogs is of key importance for diagnostics of  bursting events, detecting and characterizing  collective  behaviors, and identifying  involved users.

In this work we study the dialogs-based social networks which represent dynamical structures situated on the underlying friendship  network in \texttt{MySpace}. They involve only certain number (or type?) of the active users and their structure can vary in time, depending on the events and time window of interest.  
 We consider two sets of data with the dialogs among  users in \texttt{MySpace} social network and analyse them as complex dynamical systems  in order to determine quantitative measures of users collective behaviors. For this purpose we first develop a procedure to compile the data that have  required structure and high temporal resolution for this type of analysis. 
The datasets are then studied by methods of graph theory and statistical physics to determine topology of the dialogs-based networks and to define and compute  several other quantitative measures of the collective behaviors, in particular, the temporal correlations and the patterns of user's activity,  that can be inferred from such data. 
Furthermore, using the emotion classifier SentiStrength  \cite{paltoglou2010,thelwall2010}, which is developed  for graded estimte of the emotional content in  short informal texts,  we assess the emotion valence in  the text of each message in our datasets. This enables us to further analyse how the flow of emotion  adheres with the dialogs-based network topology and with the observed collective behaviors of  users.

\section{Empirical data from \texttt{MySpace} social network}
For the goals that we pursue in this paper high resolution data are necessary, in particular,
 (a) a {\it connected network} of users identified by their IDs as nodes, and the exchanged messages as links, and  (b) {\it each message identified} by its source and target nodes (as user IDs), time when the message is generated, and its  textual content. Arguably,  an ideal online social network for this analysis is \texttt{MySpace}, having  the data of such  structure   systematically recorded until 2010.

\subsection{Data crawling \& structure\label{sec-data}}
Communications between users in the online social networks as \texttt{MySpace} occur along friendship links: writing messages on  own {\it wall}, where they can be seen by the linked friends, or writing to the linked friend's wall directly.  
Privacy policy vary from user to user, hence some users do not allow access to their data. However, the messages that they sent to the linked users who allow the access, can be identified.

\begin{figure}[h]
\centering
\begin{tabular}{cc}
\resizebox{24.8pc}{!}{\includegraphics{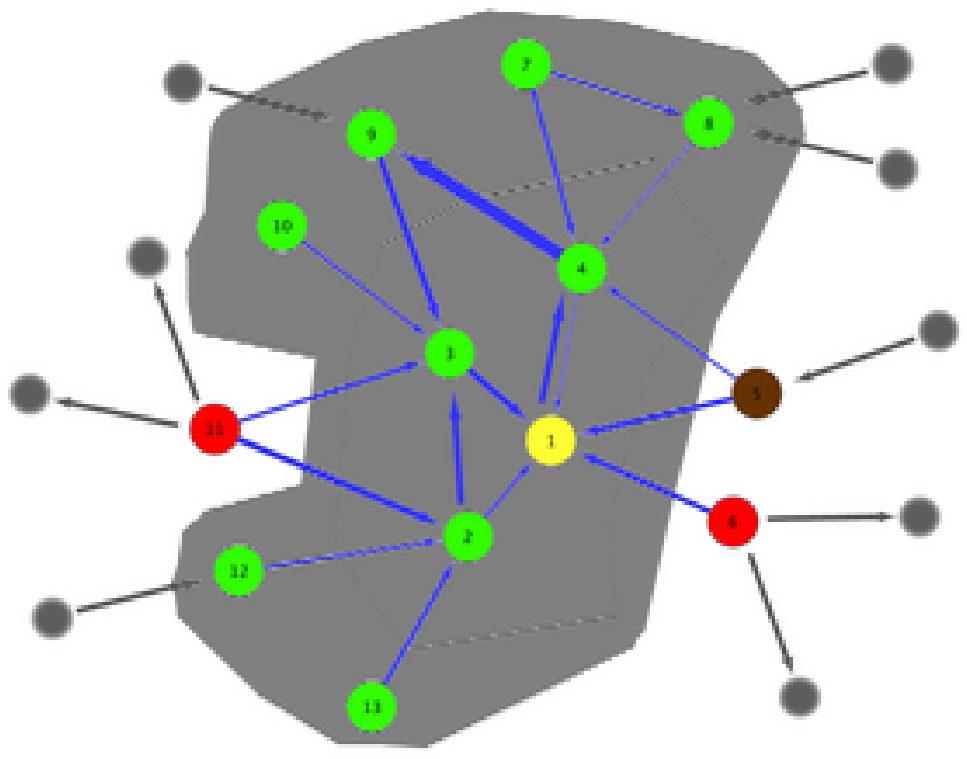}}\\
\end{tabular}
\caption{Schematic view of the parametrized breadth-first search of the dialogs occurring within specified time window, starting from a given user in \texttt{MySpace}, marked as node ``1''.  Red nodes ``6'' and ``11'' are examples of users who do not allow public access to their ``walls'', however, their messages  left on the neighboring ``walls'' can be identified. These nodes and their links are not included in the data. The presence of bots, companies, or another non-ordinary users is identified,  an example depicted by the node ``5'', and dropped from further search. 
} 
\label{fig-FBMrobot}
\end{figure}

To obtain the networked dialogs in \texttt{MySpace}, we developed a parametrized Web crawler who visits mutually linked  users and collects the dialogs (messages) which were posted by them {\it within a specified time window}.
To start  we first specify the time window that we are interested in, and  find an appropriate node, representing a {\it user} who was active within that time window.
 The crawler then proceeds to search in the neighborhood of that node in a {\it breadth-first manner}, as schematically shown in Fig.\ \ref{fig-FBMrobot},  and collecting all messages posted within that time window on the current user's wall and identifying their source and posting time. 
The links in Fig.\ \ref{fig-FBMrobot} indicate that at least one message were exchanged between the two users within the specified time window.
 Starting from the initial node marked as ``1'' the list of first layer of the connected nodes is explored,  then the search is continued from each node on that list, thus making the second layer nodes. Then the lists are swapped, and so on. 
The crawler is instructed to avoid the nodes whose data are marked as ``private'' and also the nodes which contain too many connections (probably representing a non-human user). The crawler can identify the messages posted from these types of nodes, that can be seen at the walls of their neighbors. Such nodes and their messages are not considered in the analysis.
 Full information about the exchanged messages along each discovered link is stored in  a database, in particular, the identity of the source and the target node, the creation time and full text of each message.  
Given the parametrized search, the crawler can stop either when no new nodes are found which satisfy the parameter criteria (time depth) or when it reaches a given diameter (number of layers), or accidentally for some unexpected reason.

Our data \cite{we-Data}  are compiled in 2009 and contain messages for two time windows---two months and three months depth, but starting from the same initial node.   
One dataset for two months period,  January and February 2008, consists of
  $N_{M}=80754$ messages exchanged between $N_{U}=36462$ users. The
  larger dataset  corresponds to time depth of three months, from 1st of June till 1st of September 2008, and contains
$N_{U}=64739$ users and $N_{M}=172127$ of their messages.

\subsection{Inference of the emotional valence from text of messages}
For further analysis in this work we performed automatic screening of the text of dialogs in our dataset to extract  the emotional valence of each message. 
We apply the Emotion classifier which is developed by Thelwall and coworkers \cite{paltoglou2010,thelwall2010,paltoglou2010b} for short text messages, which occur in \texttt{MySpace} dialogs. The classifier considers each message as a single document and can extract graded emotion valence as two numbers $(e_-,e_+)$, representing the intensity of the negative and the positive emotion content of the same message. 

According to Refs.\ \cite{thelwall2010,paltoglou2010,paltoglou2010b}, the classifier algorithm is based on two emotional dictionaries \textit{the
General Inquirer} (GI) and \textit{Linguistic Inquiry and Word Count} (LIWC). 
In each message the algorithm detects all words that
belong to the emotional dictionary and extracts their polarity and intensity.
The obtained scores are then modified with additional linguistic-based rules if 
special terms, such as negators (\textit{good} vs. \textit{bad}), intensifiers (\textit{liked} vs. \textit{liked very much}),  diminishers (\textit{excellent} vs. \textit{rather excellent}) are found in the neighborhood of that word in the area of $5$ words before and after the emotional term or the beginning or end of the sentence. Other special terms as  capitalization (`\textit{bad} vs \textit{BAD}), exclamation and emotion detection
(\textit{happy!} or  \textit{:-)}) are searched and treated similarly as intensifiers.  
 If an
intensifier (or diminisher) word is found then the absolute original emotional
value of the term is increased (or decreased) by one. For example, 
 \textit{bad} is given ($-3$) then \textit{very bad} is  ($-4$). Similarly, if \textit{good} is judged as ($+3$) then \textit{somewhat good} is ($+2$), while \textit{very good} is ($+4$).   The scores  $e_{+}$ and $e_{-}$ that the classifier returns represent  the maximum
positive and the maximum negative number for a given message. Accuracy and other relevant details can be found in the original Refs.\ \cite{thelwall2010,paltoglou2010,paltoglou2010b}.

The classifier returns two independent ratings for every message: one
for the positive  $e_{+}\in\{+1,+2,+3,+4,+5\}$ and one for the negative 
$e_{-}\in\{-1,-2,-3,-4,-5\}$ dimension. On this scale $e_-=-5$  correspond to very negative  and $e_+=+5$ to very positive  emotion valence. While  $e_-=-1$ and $e_+=+1$ indicate the absence of  negative and positive  emotion, respectively. 
Based on these automated ratings  we can construct the overall valence polarity of a particular  message. 
Specifically, all messages for which the scores are in the range ($e_-\leq -3$ and $e_+ =+1$ or $e_+= +2$) are considered as carrying {\it negative emotion valence}, and symmetrically the messages with ($e_+\geq +3$ and  $e_-=-1$ or $e_-=-2$) are classified as carrying {\it positive emotion valence}. Notice that this excludes certain number of messages for which the two scores are exactly balanced, $(e_-=e_+)$, although they may contain ``emotional'' words. Also the messages for which the negative and positive scores are simultaneously very high, ($e_{+}\geq3$ and $e_{-}\leq-3$), are disregarded as possible artifact of the graded-strength classifier.

\subsection{Methodology for  data analysis}
The data set for a given time window is mapped onto dialogs-based network in the following way: Each user is represented by a node and a link is inserted between the pair of nodes $(i,j)$ if a dialog (at least one message) was detected between this pair of users  within that time window. The direction of the link indicates the message from source-to-target and the link multiplicity  (weight) represents the number of messages. In some figures colors of the links are used to suggest net emotion balance along the link within the specified time window, which is derived form the original scores of the messages along that link.

Our high-resolution datasets contain valuable information which enables us to study  other features, in particular the temporal sequences of user activities and the patterns of emotion flow via  message exchange.
To study such diverse aspects of the online social networks,  appropriate  methodologies and mathematical approaches are applied. In particular:
\begin{itemize} 
\item {\it The network topology} is analysed using the graph theory methods \cite{bb-graphtheory}. Specifically, we determine several topology measures at global network level (distributions of degree, strength, weight, betweenness) and at local node-neighborhood level (assortativity measures and tests of social weak-tie hypothesis), as well as the mesoscopic level (community structure). 
For the community structure analysis in  compact and relatively small networks we  apply the eigenvalue spectral analysis of the Laplacian operator related with the  {\it weighted symmetrical} adjacency matrix \cite{mitrovic2009}.  In the case of large graphs the communities are detected by \texttt{Gephi} software, which utilizes maximum modularity approach \cite{blondel2008}.  
The maximum-flow spanning trees of our networks are determined using  greedy algorithm.

\item  {\it Temporal correlations} are studied by time series analysis. Three types of time series are extracted from the data, i.e.,  the time series of the number of messages,   and the number of messages carrying  positive or negative emotion, per  small time bin $t_{bin}=5$ minutes. 

\item {\it Patterns of user activity} are mapped directly from the dataset in the original temporal resolution.  Each action (message) of a given user at a given time is represented by a point on the user temporal pattern.
The interactivity time  is identified as the distance between two consecutive points along the time axis from the activity pattern of each user.
\end{itemize}

 Various histograms obtained in this analysis are fitted to the corresponding theoretical expressions using  either $\chi ^2$ or Maximum Likelihood Estimator (MLE)  method. 
All fits passed the $\chi ^2$--test of goodness. Logarithmic binning with the  base $b=1.1$ is often applied  to obtain smooth curves.

\section{Topology  of the Dialogs-Based  Networks in \texttt{MySpace}\label{sec-network}}

\begin{figure}
\centering
\begin{tabular}{cc}
\resizebox{24.8pc}{!}{\includegraphics{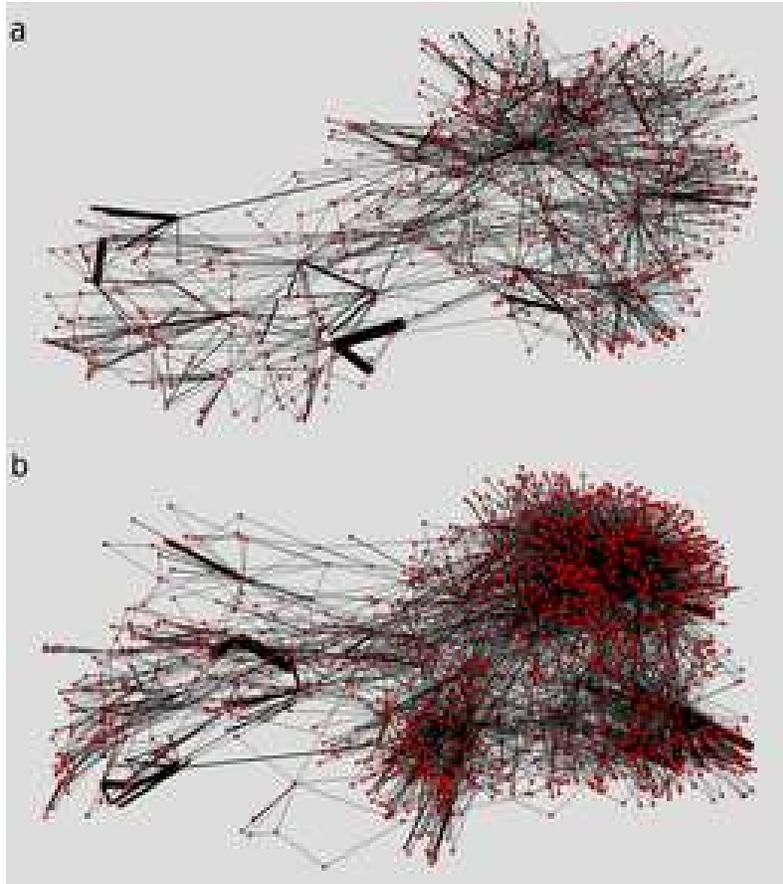}}\\
\end{tabular}
\caption{Initial part of the network of users connected by dialogs in \texttt{MySpace}, as compiled by our crawler for the time window of two months (a) and three months (b) starting from the same user-node.  
} 
\label{fig-msnet-original}
\end{figure}

Using the above described procedure of data mapping, we obtain  networks of \texttt{MySpace} users, as nodes, connected by directed weighted links, representing the messages sent from one user to another. The weight of a link indicates the number of messages along that link within the considered time window. 
In Figure\ \ref{fig-msnet-original} two examples of such dialog-based networks are shown. In the top panel a part of the network of dialogs observed within the time window  $T_{WIN}=2$ months is shown. While the lower panel shows the corresponding network for the situation  when the searched time depth is  $T_{WIN}=3$ months starting from the same initial node, as explained in section\ \ref{sec-data}.  Shown are small initial part of the corresponding dataset.
The increased time depth manifests in that larger number of nodes are connected to the network, the links density is increased as well as  the widths of some already existing links.  Moreover, the community structure---visually identified as groups of nodes, is evolving.

\begin{figure}[h]
\centering
\begin{tabular}{cc}
\resizebox{24.8pc}{!}{\includegraphics{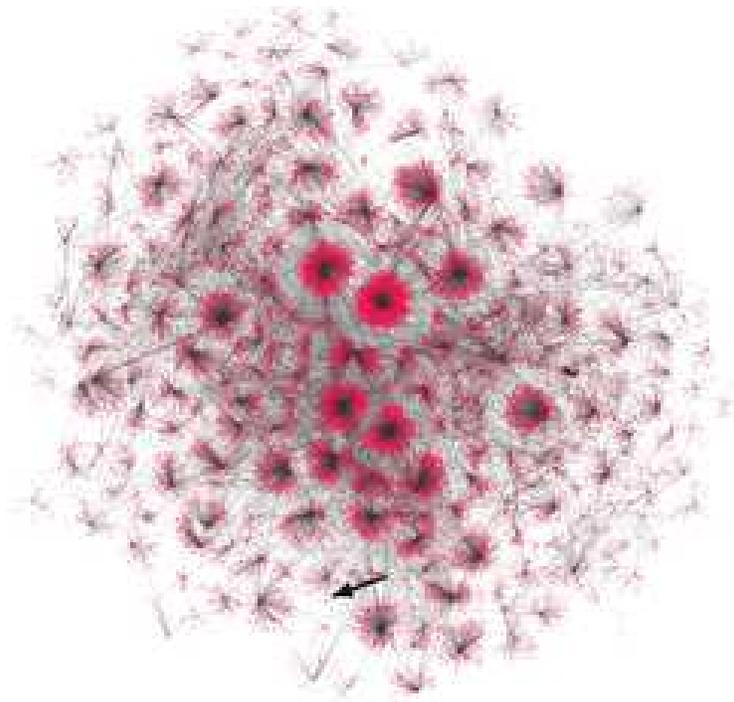}}\\
\end{tabular}
\caption{View of the network \texttt{Net2M} from the dataset of dialogs with two months depth in \texttt{MySpace}. $N_U=33649$ nodes are organized in 87 communities, seen  as blobs of different sizes.   
} 
\label{fig-2m_net_whole}
\end{figure}

\textit{Density and Reciprocity of links.}
Here we consider topology  of two networks representing all \textit{emotion classified dialogs} in 2 month time window (\texttt{Net2M}), and in 3 months time window (\texttt{Net3M}). They contain $N=33649$ and $58957$ users, respectively. These networks appear to be very sparse, cf.\ Fig.\ \ref{fig-2m_net_whole}. The average link density, defined as $\rho \equiv L/N(N-1)$, where $L$ is the number of all directed links,  is found as $\rho =3.345\times 10^{-5}$ for \texttt{Net2M}, and  $\rho =2.08\times 10^{-5}$ for \texttt{Net3M} network. Furthermore, we compute the \textit{link reciprocity}, which is defined \cite{reciprocity} as  $r\equiv (L^{\leftrightarrow}/L -\rho)/(1-\rho)$, where $L^{\leftrightarrow}$ is the number of links occurring in both directions $i\to j$ and $j\to i$ disregarding the weight. We find  $r=0.0227$ for \texttt{Net2M}, and $r=0.0214$ for \texttt{Net3M} network, i.e., the reciprocity is barely positive. According to \cite{reciprocity}, in social networks larger positive reciprocity is expected. The clustering coefficient $Cc=$0.013 and 0.014 are found for these two networks when the directedness of the links is ignored.
We also consider a reduced network, termed \texttt{Net3321}, which is extracted from two months dataset. In this network the users who sent and received less than four messages within two months were excluded. Thus reduced network contains $N_U=3321$ nodes and is more compact, i.e., $\rho=7.19\times 10^{-4}$ and has larger link reciprocity $r=0.118$ and the clustering coefficient $Cc=$0.084, for directed, and 0.115 for undirected links.

\begin{figure}
\centering
\begin{tabular}{cc}
\resizebox{24.8pc}{!}{\includegraphics{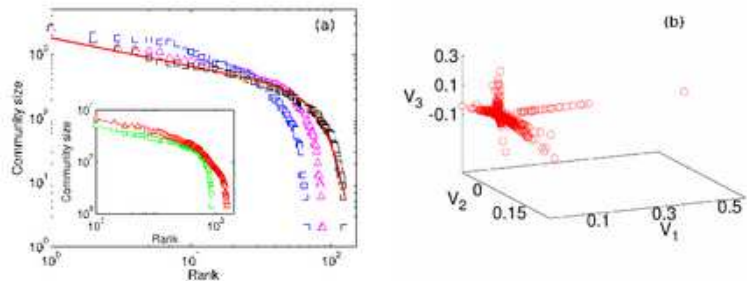}}\\
\end{tabular}
\caption{(a) Ranking distribution of the size of communities in the  networks  \texttt{Net2M} of two months dialogs in \texttt{MySpace}. Three curves from front to back are for higher, standard, and lower resolution, respectively. Fit line has the slope $0.44\pm 0.02$ and the cut-off $81.85\pm 0.78$. Inset: Same but for two and three months dialogs  networks \texttt{Net2M} ($\square$) and \texttt{Net3M} ($\triangle$) with standard resolution. (b) The presence of communities in \texttt{Net3321} indicated by the branches in 3-dimensional  plot of the eigenvectors for three lowest nonzero eigenvalues of the Laplacian.
} 
\label{fig-spectrum}
\end{figure}

\textit{Community structure.} In Fig.\ \ref{fig-2m_net_whole} the entire network of \texttt{Net2M} is shown. The network is organized in large number of small \textit{communities}, specifically  87 communities  shown in Fig.\ \ref{fig-2m_net_whole} are obtained by weighted maximum modularity algorithm \cite{blondel2008} with \texttt{Gephi} software. Largest community contains 2543 users. It is interesting to note that the size distribution of these communities obeys Zipf's law. In Fig.\ \ref{fig-spectrum}a  two curves in the inset are the ranking distributions of the community sizes which are detected in the networks of two months dialogs and three months dialogs, respectively. It should be stressed that the number of communities that one observes depends on the resolution in the algorithm. By increasing the resolution, some communities can further split into two or more groups, and oppositely,  join together to make a larger community when the resolution is decreased. However, the scale-free organization of communities up to certain size persists (with a changed slope), as shown in the main Fig.\ \ref{fig-spectrum}a. 
 
In the \texttt{Net3321}, the nodes with small strength $\ell _{in}+\ell_{out} \leq 4$ are excluded, as mentioned above. 
(The directed network  is further analysed in section\ \ref{sec-patterns}.) 
Here in Fig.\ \ref{fig-spectrum}b we confirm that this more compact network also exhibits a  community structure. We use the eigenvalues spectral analysis \cite{mitrovic2009} of the normalized Laplacian  ${\cal{L}}=\delta_{ij} -\frac{W_{ij}^S}{\sqrt{\ell_i\ell_j}}$, where $\ell_i$ and $\ell_j$   are  total strengths and $W_{ij}^S\equiv W_{ij}+W_{ji}$ symmetrical weighted adjacency matrix of the network. Property that the eigenvectors localize on subgraphs \cite{mitrovic2009} is used to identify communities.
Note that the communities detected in the dialogs-based network are dynamical, i.e., related with a particular part of the network structure that has been actually used within the considered time window.

\begin{figure}
\centering
\begin{tabular}{cc}
\resizebox{24.8pc}{!}{\includegraphics{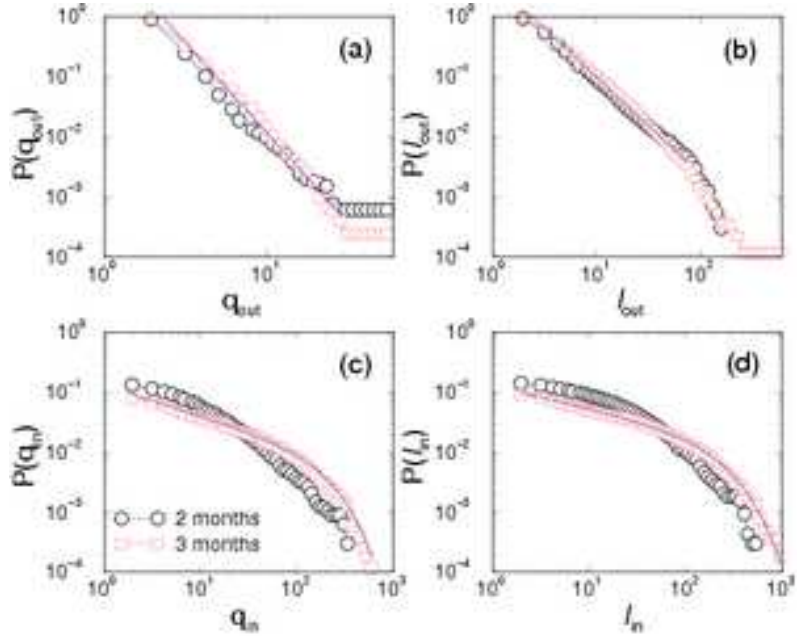}}\\
\end{tabular}
\caption{Out-degree  (a) and Out-strength (b), and In-degree (c) and In-strengths (d) of nodes  on the networks of \texttt{MySpace} dialogs within 2-months and 3-months time window.  Log-binned data. Dotted lines are fits according to Eq.\ (\ref{eq-distributions}).
} 
\label{fig-msnet-distributions}
\end{figure}

\begin{figure}
\centering
\begin{tabular}{cc}
\resizebox{24.8pc}{!}{\includegraphics{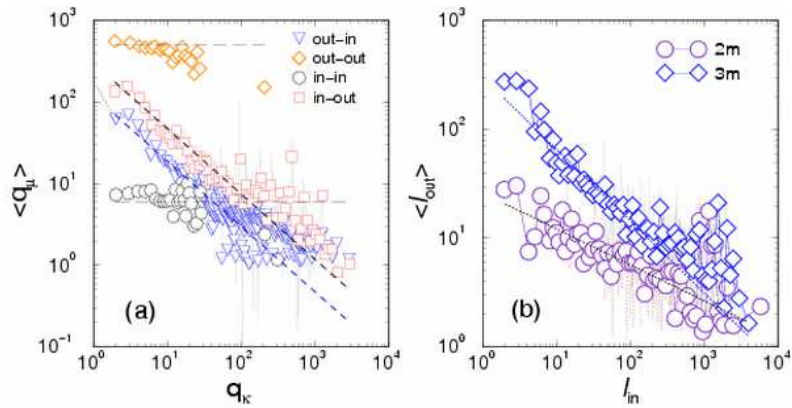}}\\
\end{tabular}
\caption{Mixing patterns on \texttt{MySpace} dialogs network: (a) correlations between  in- and out-degree (four combinations) for the dialogs  within 3 months time window, and (b) in- and out-strengths for 3 months and 2 months time window. Log-binned data. Dashed  lines indicate  slope $\mu \sim 1$, while  two dotted lines are  for $\mu=0.33$ and $0.86$. 
} 
\label{fig-msnet-mixing}
\end{figure}

{\it Nodes inhomogeneity \& mixing patterns.}
The  node's degree and strength distributions and mixing patterns for the networks \texttt{Net2M} and \texttt{Net3M} are are computed and the results are shown in Figs.\ \ref{fig-msnet-distributions} and \ref{fig-msnet-mixing}. 
Both degree and strengths distributions can be fitted by the mathematical expression
\begin{equation}
P_\kappa(X) = B_\kappa X^{-\tau_\kappa}e^{-X/X_{0\kappa}} 
\label{eq-distributions}
\end{equation}
where the exponent $\tau_{\kappa}$ and the characteristic cut-off length $X_{0\kappa}$ may vary, depending on the type of the links ($\kappa =$in, out) and the time-window where the dialogs are observed. Specifically, the distributions of
out-link degree and out-link strengths follow a power-law decay. The exponent $\tau_{out}=3.01\pm 0.07$ before the cut-off $X_0=18$ for the degree, and $\tau_{out}=1.62\pm 0.02$ and $X_0=98$ for the strength distributions are found. In Fig.\ \ref{fig-msnet-distributions}a,b fitted are only parts with the power-law dependence. These distributions appear to be quite stable with respect to  time-depth, i.e.,  similar slopes are found  for the distributions from 2-months and 3-months time window data. 
On the other hand, the in-links degree and in-links strengths receive the form (\ref{eq-distributions}) only when the time-depth is large enough. They are characterized by much smaller exponent $\tau_{in} <1$ and large cut-off: $\tau_{in}=0.53\pm 0.07$, $X_0=226\pm 32$ for the in-degree, and $\tau_{in}=0.88\pm 0.04$, $X_0=248\pm 17$ for the in-strength. The best fit, as shown in Fig.\ \ref{fig-msnet-distributions}, is received when a small stretching exponent is added, 1.25 for the in-degree and 1.06 for the in-strength distributions. 
While the out-degree and out-strength are  controlled by the node itself---representing the user's actions directed to its neighbors, the in-degree contains the cumulative action of  all neighbors directed to that node's wall. Hence, the lower exponent is expected as well as  the large cut-off, which reflects  the diversity of first-neighborhoods of nodes on the network.

The assortativity measures are another characteristics of the network's  topology at the level of node's neighborhood.
Specifically, the situations when the nodes with large degree are linked to each other (assortativity) or oppositely, the nodes with large degree have a large number of nodes with small degree (dis-assortativity), will be expressed by increasing (decreasing) slopes in the degree-correlations plots, like the ones in Fig.\ \ref{fig-msnet-mixing}a. In view of the \textit{weighted and directed} nature of the dialogs network, we can determine several such measures. The results are shown in Fig.\ \ref{fig-msnet-mixing}a,b for degree and for strength mixing, respectively. Specifically, considering a node with a given in-degree, plotted  along the x-axis $q_{\kappa=in}$, and  computing the average out-degree of the nodes linked to it, $<q_{\mu=out}>_{nn}$, referred as $in-out$ in the Legend, we find a dis-assortative pattern, i.e., $<q_{\mu=out}>_{nn} \sim q_{in}^{-\mu}$ with the decreasing slope $\mu \sim 1$. The tendency to flattening in the neighborhood of large-degree nodes suggests the occurrence of communities.  Similar finding applies for the $out-in$ mixing. On the other hand, no assortativity measures can be observed in  $out-out$ and $in-in$ patterns, represented by flat curves in   Fig.\ \ref{fig-msnet-mixing}a.   The results are from the 3-months dialogs dataset. The disassortative measures are already present in smaller-depth dialogs, although with a different negative slope. In Fig.\ \ref{fig-msnet-mixing}b the results for the $in-out$ strength mixing are compared for 2-months and 3-months dialogs networks. 
These findings suggest that a particular pattern with  a large number of small-degree nodes communicating  with one large-degree node occurs very often. 
This dynamical pattern, based on the friendship connections in \texttt{MySpace} which already shows some tendency towards disassortative mixing \cite{myspace2007}, is in clear contrast with the assortativity in  common social structures  \cite{newman2003} and  static friendship networks in \texttt{Facebook} \cite{facebook2011}. 

\begin{figure}
\centering
\begin{tabular}{cc}
\resizebox{24.8pc}{!}{\includegraphics{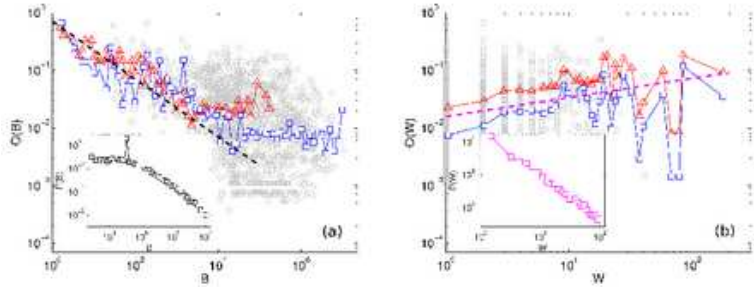}}\\
\end{tabular}
\caption{Averaged overlap as function of betweenness centrality $B$ (a) and of weight $W$ (b) for symmetrical links on \texttt{MySpace} dialog network of 2 months time window ($\square $) and \texttt{Net3321} ($\triangle $). Dashed lines in the left and the right panel indicate slopes $-1/2$ and $+1/3$, respectively,  conjectured in Ref.\ \cite{ST12}. Insets: Distributions  $P(B)$ of betweenness and $P(W)$ of weights for the network of 2 months depth. Data are logarithmically binned. 
} 
\label{fig-msnet-overlap}
\end{figure}

\textit{Confirmation of  the weak-ties hypothesis.} For quantitative measures of traditional social dynamics weak-ties hypothesis, it is widely accepted to determine correlations in \textit{betweenness} centrality $B_{ij}$ of a link and \textit{overlap} $O_{ij}$ of two neighboring  nodes on the network \cite{onela2007,ST12}. Specifically, for the link $(ij)$, betweenness $B_{ij}=\sum_{s,t\neq s}\frac{\sigma_{st}(ij)}{\sigma_{st}}$ is given by the fraction of shortest paths $\sigma_{st}(ij)$ between pair of nodes $(s,t)$  that pass through that link, compared to all shortest paths between $(s,t)$ and averaged over all pairs on the network \cite{bb-graphtheory}.  
Note that for this purpose the network is considered as undirected! 
Overlap of two adjacent nodes $i$ and $j$ is computed as $O_{ij}= \frac{m_{ij}}{q_i + q_j -2 -m_{ij}}$, where $q_i$ and $q_j$ stands for total degree of nodes $i$ and $j$ and $m_{ij}$ is the number of nodes that are common neighbors to both of them. 
In traditional social dynamics it is expected that the overlap  increases with bond strength, i.e., $O_{ij}(W) \sim W^{\eta_1}$. Moreover, due to the ubiquitous community structure, the  bonds with large betweenness, e.g., connecting different communities, should not have large overlap. Consequently, $O_{ij}(B)\sim B^{-\eta_2}$. In the networks of online social contacts weak-ties hypothesis was confirmed in e-mail networks \cite{onela2007} and online games \cite{ST12}. It was conjectured in \cite{ST12} that universal exponents $\eta_1=1/3$ and $\eta_2=1/2$ should apply. 
Our results for \texttt{MySpace} two-months dialogs network and for more compact \texttt{Net3321}, shown in Fig.\ \ref{fig-msnet-overlap}a,b,  confirm this conjecture of Ref.\ \cite{ST12}. In the insets we also show computed histograms of the betweenness $P(B)$ and of the weight $P(W)$ for all links in two-months dialogs window. Power-law tails in these distributions suggests diversity in both the organization of the communities and in the intensity of communications inside these communities.

\section{Activity Patterns  and Emotion  Flow\label{sec-patterns}}

\begin{figure}[!h]
\centering
\begin{tabular}{cc}
\resizebox{24.8pc}{!}{\includegraphics{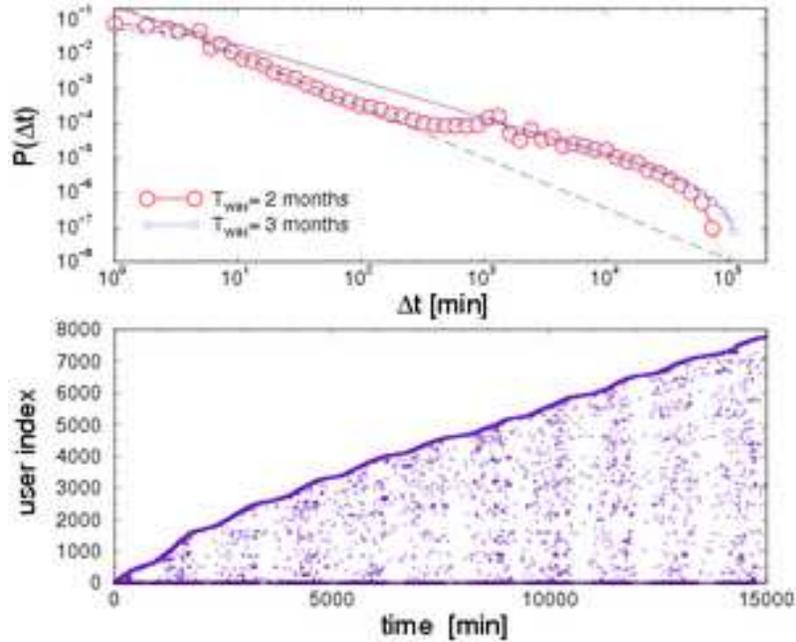}}\\
\end{tabular}
\caption{(bottom) Example of user-activity pattern from  \texttt{MySpace} dialogs. Shown are first 10 days from the 2-months time window data. (top) Distribution  $P(\Delta t)$ of the user delay time $\Delta t$, averaged over all users. Two curves are  for  the 2-months and 3-months time window datasets. Data are logarithmically binned. Dashed and dotted lines are explained in the text.
} 
\label{fig-uspattern}
\end{figure}

{\it Temporal patterns of  user activity}. 
In our high-resolution data information about every user activity over time
can be presented as a pattern, an example is shown in lower panel in Fig.\ \ref{fig-uspattern}. Time in the original resolution is plotted along x-axis, and  each integer number along y-axis stands for an user index. The indexes are ordered by user's first appearance in the dataset.

Two characteristic features of these temporal patterns are (cf.\ Figs.\ \ref{fig-uspattern}):
\begin{itemize}
\item {\it Arrival of new users}, depicted with the top boundary of the pattern, follows daily cycles. With the appearance of new users (relative to the beginning of the dataset) the system experiences increased activity, manifested as larger density of points below each ``wave'' of new users. Note also the stripes inclined upwards, which indicate possible correlated actions of the users involved in later times (in section \ref{sec-avalanches} the temporal correlations are studied in detail).  Both the arrival of new users and the increased activity of all other users obey periodicity, compatible with the {\it circadian cycles}, which is carried  over from user's offline life. Similar features are observed  in Blogs and Diggs \cite{mitrovic2011} and other social systems \cite{malmgren2009}, confirming the importance of  circadian cycles in the online dynamics. 

\item {\it Delay (interactivity) times of user actions} $\Delta t$, defined as time between two consecutive actions of a user, is a quantitative measure of fractality of the temporal activity  pattern. Namely,
following the line of a given user we find no characteristic distance between two consecutive  activity points. The distribution of distances $\Delta t$  between subsequent points for a given user, then averaged over all users in the dataset, is shown in Fig.\ \ref{fig-uspattern} top panel. The broad distribution $P(\Delta t)$ shows faster decay, approximately as $\sim (\Delta t)^{-1.475\pm 0.099}$,  for the short delays in the range from $[5-85]$ minutes, the slope is indicated by dashed line. Whereas, a majority of the delay times appear to be one day, corresponding to the peak in the middle, or longer. These long delays can be fitted with the expression $P(\Delta t) =B(\Delta t)^{-\tau _{\Delta}}exp\{-\left(\Delta t/\Delta _0\right)^\sigma\}$, with the parameters $\tau_\Delta = 1.061\pm 0.009 $ and 
and $\Delta_0=75600$ and $\sigma =2$.  
These parameters are for the 3-months window dataset, the fitted curve is shown by dotted line in Fig.\ \ref{fig-uspattern} top panel. Same expression with a similar exponent, the cut-off $\Delta_0=52000$ and the stretching $\sigma=3$ fits the dataset from 2-months time window, shown  in Fig.\ \ref{fig-uspattern} by dotted curve.
\end{itemize}

Occurrence of the power-law decay  in the delay-time distribution was recently observed in different types of data related to online dynamics \cite{castellano2009,mitrovic2010a,vazquez2006}. Theoretical arguments of the random queuing processes have been used to derive the universal scaling exponent 3/2 in the delay-time distributions  \cite{grinstein2008}.    
The present analysis of the data from \texttt{MySpace} social network suggests the exponent close to 3/2 only for the short delay-times (between 5 and 85 minutes). However, the one day or longer delays are more probable with the exponent $\tau_\Delta \gtrsim 1$, suggesting another possible  mechanism.  It is also interesting to note that delays shorter than 5 minutes are equally probable, which suggests  $t_{bin}=5$ minutes as  natural temporal resolution for these type of processes.

\begin{figure}
\centering
\begin{tabular}{cc}
\resizebox{24.8pc}{!}{\includegraphics{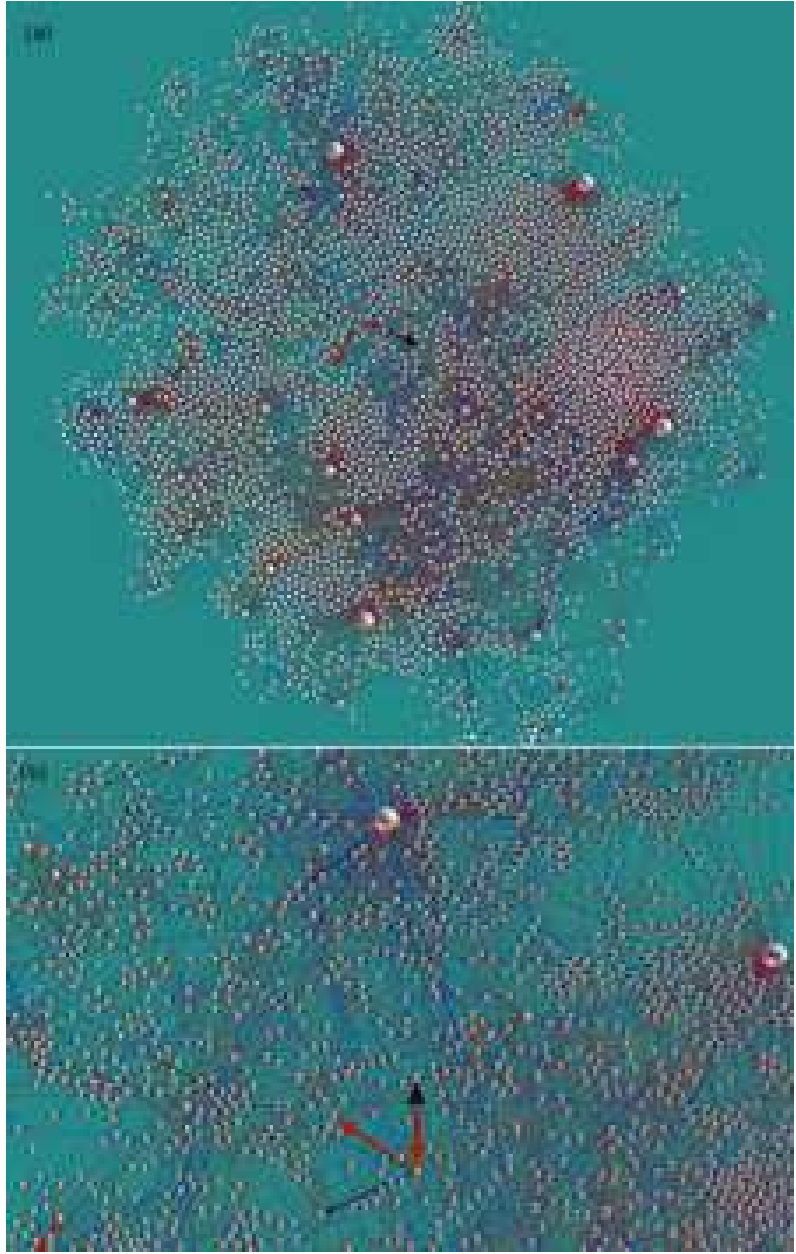}}\\
\end{tabular}
\caption{\texttt{MySpace} dialogs network \texttt{Net3321} (a) and a zoomed part of it (b). Size of nodes corresponds to node's degree centrality. The width of links represents the cumulative number of messages within two months period, while the color indicates their emotion valence---positive (red), neutral (blue) and negative (black). 
} 
\label{fig-Net_positive}
\end{figure}

{\it Structure  of the emotional dialogs.}  We consider in detail the emotion content transmitted along the links in the network \texttt{Net3321}. As stated above, the directed weighted link $W_{ij}$ on this network represents the number of messages sent from the node $i$ to the node $j$ within the time window of two months. Here we also include the emotion valence of these messages. By summing up the emotion contained in each message along the link we obtain the overall valence of the link, i.e.,  as positive,  negative or neutral link. The network is shown in Fig.\ \ref{fig-Net_positive}a with the links colored according to their emotion content---red (positive), black (negative) and blue (neutral).  The size of the nodes indicates their degree centrality on the network. Each node carries a label---unique user ID index from the original data. The zoomed part of the network, which is  displayed in  Fig.\ \ref{fig-Net_positive}b, illustrates a typical structure of the emotion-carrying links between hubs and the number of  small-centrality nodes surrounding them. 

\begin{figure}
\centering
\begin{tabular}{cc}
\resizebox{24.8pc}{!}{\includegraphics{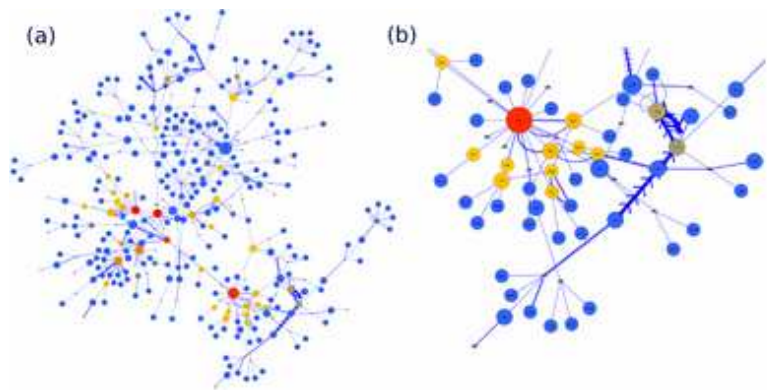}}\\
\end{tabular}
\caption{Largest connected component of the subnetwork with negative dialogs (a) and a its zoomed part  (b), on which directions and weights of the negative valence dialogs are indicated. Nodes are marked by different size comparable with the out-degree and color determined by the betweenness-centrality of the nodes on the entire network \texttt{Net3321}, from which the negative links subnetwork is extracted.
} 
\label{fig-Net_negative}
\end{figure}

Considering the emotion contents of the messages along the links, we find that the positive emotion (attractiveness) dominates the connections in \texttt{MySpace} dialogs. Whereas, the links carrying negative emotion (aversiveness) occur rather sporadically, acting as ``impurity'' in the sea of positively charged contacts.  The temporal correlations of the positive (negative) emotion messages are further studied in section\ \ref{sec-avalanches}. Here we analyse the topology of the emotional (and otherwise important) connections on the network.
 We extract the subnetwork of negative links, whose fragments are found in different parts of the \texttt{Net3321}. The largest connected component of the negatively linked network  is shown in Fig.\ \ref{fig-Net_negative}a.  
Enlarged part of this network, shown in Fig.\ \ref{fig-Net_negative}b, demonstrates typical flow patterns of the negative-emotion messages. Specifically, a node  may act as a source or a sink of the negative links, can transmit or disseminate the emotion, or be involved in multiple reply-to events with the same emotion valence. 
Note that these two subnetworks with  positive and negative emotion  links are integrated into each other, and  also that some messages (links) can be considered as neutral, i.e., carrying information, but not emotion. Therefore, 
the nodes that  change the valence of the emotion messages exist and have a special role as pinning centers for the propagation of the emotional bursts on the network.

\begin{figure}[!h]
\centering
\begin{tabular}{cc}
\resizebox{24.8pc}{!}{\includegraphics{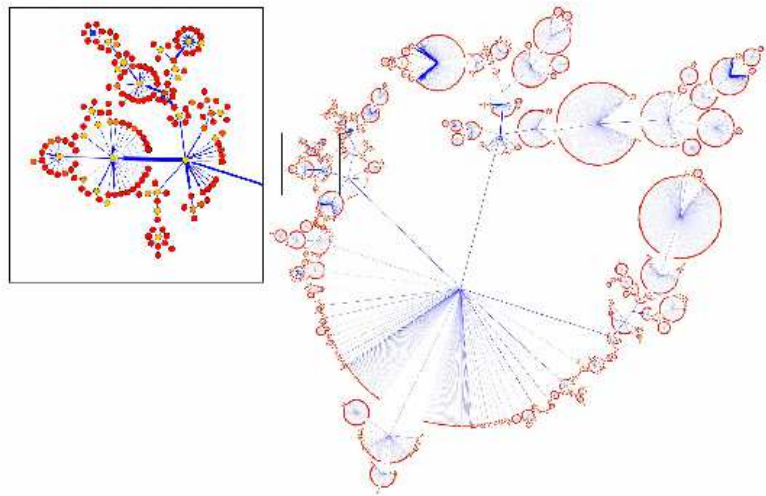}}\\
\end{tabular}
\caption{Maximum-flow spanning tree of the  network  \texttt{Net3321}. Enlarged part on the left shows a typical structure away from the root. 
} 
\label{fig-MSFT}
\end{figure}

For most of the links  the computed overlap is related with the widths of links, cf.\ Fig.\ \ref{fig-msnet-overlap} according to the social tie hypothesis. However, sometimes strong links, i.e., carrying large number of messages   appear (visible on the network in 
Fig.\ \ref{fig-Net_positive})  dis-proportionally to the topological centrality of the adjacent nodes. 
To find how the most of  messages (and emotion) flow on the entire  network, we analyse the maximum-flow spanning trees. These are suitable representation of the network where each node is connected to the tree by its \textit{strongest link}.  
In Fig.\ \ref{fig-MSFT} the maximum-flow spanning three of the \texttt{Net3321} is shown. 
The tree is constructed using a variant of greedy algorithm by ordering the \textit{total weights} of the links $W_{ij}+W_{ji}$. It shows considerable  side branching, which suggests heterogeneity in the intensity of the dialogs inside the existing  communities. Moreover, it often occurs that a small-degree node interpolates in the branching process and transmits the flow of messages between the hubs. This feature is in agreement both with the observed community structure and the dis-assortativity of the dialogs-based network, discussed in section\ \ref{sec-network}. 
It further supports the conclusion  that the dynamical structure in the online social network \texttt{MySpace} is different from the networks of conventional social contacts.

\section{Correlations in time series with emotional messages\label{sec-avalanches}}
\begin{figure}[!h]
\centering
\begin{tabular}{cc}
\resizebox{24.8pc}{!}{\includegraphics{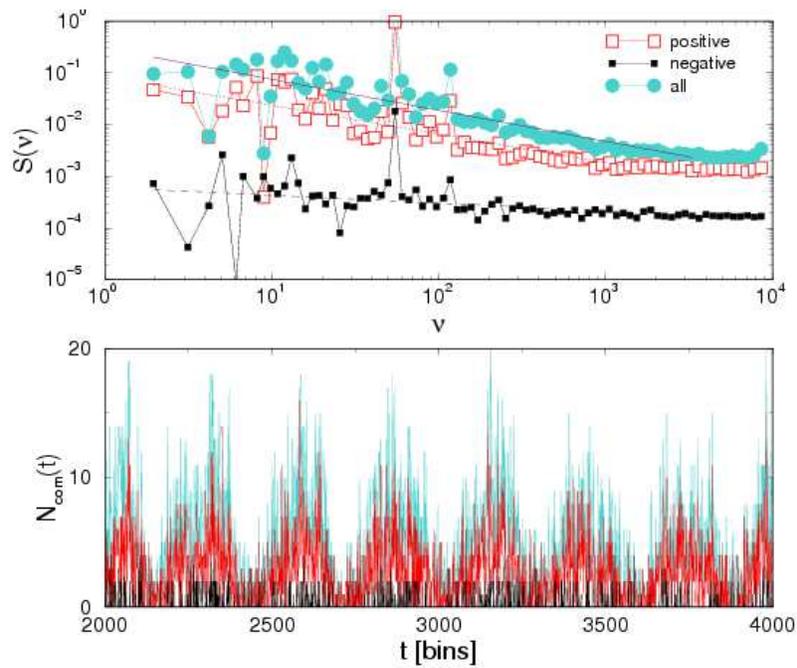}}\\
\end{tabular}
\caption{Time series of the number of messages from \texttt{MySpace} dialogs network for 2-months time window: all identified messages  (cyan), and  the messages classified as carrying positive (red) and negative (black)  emotion valence.  Time axis in bins of 5 minutes. Length of each time series is 16384 time bins. For better vision shown is only a small part corresponding to one week time span.  Top panel: power spectrum of these time series. Log-binned data. Fit lines are explained in the text.
} 
\label{fig-timeseries}
\end{figure}

The stochastic processes governing the communication with emotional messages  among the users in \texttt{MySpace} can be studied from the point of view of the time-series analysis. 
From the datasets of the networked dialogs in a given time window various  time-series are constructed here, for instance,  the series which contain the \textit{number of all identified messages per small time bin}.  Similarly, we construct  the time-series of the number of messages carrying positive/negative emotion valence.  The time bin $t_{bin}=5$ minutes is used as the natural resolution in these data. Examples  of these time-series and their power spectra are shown in Fig.\ \ref{fig-timeseries}.

The time series in Fig.\ \ref{fig-timeseries} exhibit fluctuations with strong daily periodicity (circadian cycles), observed also in the activity patterns in Fig.\ \ref{fig-uspattern} and discussed above.
This periodicity is manifested as  a pronounced peak in the power spectrum at the corresponding frequency, i.e,  at the index value $\nu \approx 56$ in this case. 
Apart from the peak, the power spectra in Fig.\ \ref{fig-timeseries} are of the colored-noise type (fractal time series). Specifically, the spectrum can be expressed as  $S(\nu)\sim 1/\nu^\phi$ for a range of frequency index below approximately 3000 (corresponding to the time scale longer than  2 hours) in the case of time series of all messages. Similar feature is found in the time series of messages with positive emotion for  $\nu < 1000$ (or $t>5$ hours) approximately. 
These features of the time series suggest occurrence of the long-range correlations in the fluctuations in number of messages of all types and in the messages with positive-emotion valence.  
Whereas the spectrum of the negative-valence messages appears to be much closer to the white noise signal ($\phi\gtrsim 0$). The corresponding exponents are: 
$\phi^{a}=0.59 \pm 0.08$, $\phi^{+}= 0.55 \pm 0.08$, and $\phi^{-}=0.15 \pm 0.06$ where symbols $a,+,-$ stand for all messages, and messages with positive and negative emotion valence, respectively.
For completeness, we have also computed the Hurst exponent $H$,
 which measures the strength of fluctuations of these time series. In particular, for the time series of length $T$, it is determined  from the power-law segment in the dependence  $D(n)/\sigma (n) \sim n^H$, where 
 $D(n)$ is the maximal deviations  of the cumulative time series $\sum_k^n(Y(k)-<Y>)$ and $\sigma (n)$ its standard deviations in a time window of varying length $n=1,2,\cdots T$. We find the following values $H^{a}= 0.82 \pm 0.03$, $H^{+}= 0.83 \pm 0.04$, and  $H^{-}=0.62 \pm 0.03$ for these three time series. It is interesting to note that the Hurst exponents for all time series are larger than 1/2,  suggesting the \textit{persistent fluctuations} in the overall activity and in the emotional messages of both polarity. Moreover, the scaling relation $\phi^a=2H^a -1$ is satisfied (within numerical error bars).
Further study of  the mechanisms behind these fractal time series and  the avalanches of emotional comments is left out of this work \cite{we-avalanches}.

\section{Conclusions\label{sec-conclusions}}

We have performed a comprehensive anlaysis of the empirical  data of user dialogs  from the social network  \texttt{MySpace} focusing  on the quantitative analysis of users collective behavior.   Our methodology, that can be used across large class of online systems with user-to-user interactions and high-resolution data, includes: compiling the data of suitable structure,  extracting emotion content from texts of messages by automated methods tailored for this type of textual documents, and analysing the data with graph theory and statistical physics by properly accounting  for the nonlinear dynamic effects. 
Our main findings can be summarized as follows:
\begin{itemize}
\item {\it Dialogs-based networks} in \texttt{MySpace} are dynamical structures built upon ``friendship'' connections. They organize in a large number of communities of various sizes and  the ``weak-tie'' hypothesis holds in a manner similar to online games and e-mail networks. The actual use of  links (within a given time window) reveals unbalanced message flow, yielding different organization of in-coming and out-going links; several hubs emerge to whom communication is directed from a large number of  ``small'' nodes, manifested in  strongly dis-assortative mixing; 
Furthermore, the emotion content of the messages passed along these links, averaged over the entire time window, is dominated by positive emotion (attractiveness), while the links with negative emotion  (aversiveness) appears less often, but make a specific local pattern on the network.

\item {\it Self-organized processes} of message  exchange among the users have long-range temporal correlations of various degrees and persistent fluctuations,  which  clearly depend on the emotion content of the messages. 

\item {\it Robust patterns of user behaviors} are observed, which are linked with circadian cycles.  Power-law distribution of delay-times with a small exponent for the delays longer than one day is found, suggesting that mechanisms different from queuing of tasks may be responsible. 
\end{itemize}

In conclusion, the presented multi-analysis approach sheds a new light onto the actual problem of the functional  structure in the  online social networks.
The studied patterns in \texttt{MySpace}, considered as one of the largest ``social  sites'', reveals dynamical structure that by many measures disagree with the common social networks. 
The self-organized dynamics with message exchange leads to specific organization of users where large diversity of groups is found as well as the importance of individuals within these groups. Moreover, the emerging collective behaviors depend on both intensity of communications and amount of emotion passed with the messages. 
Disproportional dominance of  positive emotions (attractiveness) may also suggest presence of  euphoria ( ``24 hours party''-like behavior). 
 Our quantitative analysis with the results presented in sections\ \ref{sec-network}, \ref{sec-patterns}, and \ref{sec-avalanches} can help in the ongoing social and psychology research on the problems ``who'' uses the social networks? and ``how'' the online social networks are used? and serve as a basis for further research and theoretical modeling.

{\bf Acknowledgments:}
{Funding for this research was received in parts from the  program P1-0044 of the Research agency of the Republic of Slovenia, the project no. P1-0044-3. 
M.\v S. also thanks the research projects ON171037 and III41011 of the Republic of Serbia and the  project from FP7-ICT-2008-3 under grant agreement n$^o$ 231323.
  We would like to thank George Paltoglou for providing  user-friendly version of the emotion classifier described in Ref.\ \cite{paltoglou2010}.}

{\bf Autror's Contribution:} Designed the research: BT; Contributed new software  and compiled the data: M\v S; Performed  emotion classification of the data: MM; Analysed the results and produced figures: BT, VG, M\v S, MM;  Wrote the paper: BT.






\end{document}